\documentclass[journal]{IEEEtran}

\usepackage{amssymb}
\usepackage{amsmath}

\usepackage{amsthm,theorem}
\usepackage{amssymb}
\usepackage{graphicx,tikz}
\usetikzlibrary{arrows,automata}

\usepackage{array,cite}
\usepackage{subfigure,epsfig}

\title{\huge Design and Analysis of Wireless Communication Systems Using Diffusion-Based Molecular Communication Among Bacteria}
 \author{
 	Arash Einolghozati,~\IEEEmembership{Student Member,~IEEE,}
	Mohsen Sardari,~\IEEEmembership{Student Member,~IEEE, and}\\
 	Faramarz Fekri,~\IEEEmembership{Senior Member,~IEEE}
 \thanks{Arash Einolghozati, Mohsen Sardari, and  Faramarz Fekri are with the School of Electrical and Computer Engineering, Georgia Institute of Technology, Atlanta, GA, 30332 USA. (e-mail: einolghozati@ece.gatech.edu; mohsen.sardari@ece.gatech.edu; fekri@ece.gatech.edu). 

This material is based upon work supported by the National Science Foundation under Grant No. CNS-111094. The content of this paper has been presented in part in~\cite{ISIT2011_Arash, ITW2011_Arash,INFOCOM2012_Arash}.}
 }

\begin{document}
\maketitle

\begin{abstract}
The design of biologically-inspired wireless communication systems using bacteria as the basic element of the system is initially motivated by a phenomenon called \emph{Quorum Sensing}. Due to high randomness in the individual behavior of a bacterium, reliable communication between two bacteria is almost impossible. Therefore, we have recently proposed that a population of bacteria in a cluster is considered as a bio node in the network capable of molecular transmission and reception. This proposition enables us to form a reliable bio node out of many unreliable bacteria. 
 In this paper, we study the communication between two nodes in such a network where information is encoded in the concentration of molecules by the transmitter. The molecules produced by the bacteria in the transmitter node propagate through the diffusion channel. Then, the concentration of molecules is sensed by the bacteria population in the receiver node which would decode the information and output light or fluorescent as a result. The uncertainty in the communication is caused by all three components of communication, i.e., transmission, propagation and reception. We study the theoretical limits of the information transfer rate in the presence of such uncertainties. Finally, we consider M-ary signaling  schemes and study their achievable rates and corresponding error probabilities. 
\end{abstract}

\section{Introduction}
\label{sec:introduction}

The use of molecular signaling as a means of communication is inspired from naturally occurring communication between bacteria in a process called Quorum Sensing (QS)~\cite{Bassler1999}. The QS enables bacteria to perform a collective task  which needs to be done synchronously only when the density of the bacteria exceeds a certain threshold.  Bacteria use molecules to exchange information among themselves for performing tasks otherwise impossible~\cite{kaplan1985,Bassler1999}. Some examples for these coordinated tasks are light production and attacking the host by bacteria. Molecular communication between bacteria is conducted in a chain of reactions in such a way that the population of bacteria can reliably infer information about their density in the environment.  In particular, each individual bacterium in a population releases more molecules to the environment in reaction to the concentration of molecular signals in the medium. This chain of reactions increases the density of molecules over time. The aggregate concentration of molecules in the environment (sensed by the same population of bacteria) is a measure of the local density of bacteria. Bacteria perform their task synchronously when the concentration of molecules surpasses a threshold. The output of quorum sensing process can be in various forms. For example, the bacteria can emit light or produce Green Fluorescent Protein (GFP) which can be used to convey information to the outside world. For the rest of the paper, we assume that the output of bacteria is in the form of GFP.

New applications and designs are constantly emerging from manipulation of the genetic content of bacteria (e.g., plasmid), and have this transported between different nano machines~\cite{multi_genetic}. In such approaches,  more information may be transferred in the network. However, the major drawback with the genetic content encoding approach is the high unreliability. On the plus side, the genetic content approach may lead to higher capacity than the one in the molecular communication.
 Further, QS is used in~\cite{Danino2010} to design biological clocks through the regulation of the output of a population of bacteria to alternate periodically. There has been also new research in network engineering inspired by this phenomenon. For example, models for forming a network via molecular communication are given in~\cite{INFOCOM2012_Arash,Akyildiz2011, ISIT2013_Arash,Pierobon2010,Modulation_Akyildiz}, relying on encoding of information in the variations of concentration of molecules. There are also other lines of research in communication of biological entities.  Among them are encoding the information in the \emph{timing} of emission of molecules~\cite{eckford,Eckford_Drift,rose2011} and using $Ca^{2+}$ signaling for communication~\cite{Nakano_Relay}. 

All these studies have inspired researchers to investigate further the molecular communication among bio entities both at the system level as well as the information-theoretic sense. In~\cite{Monaco}, fundamentals of molecular communication in nano-networks have been discussed. They include the channel description, channel capacity discussion, protocol design and experimental validation setup.In~\cite{Nakano_opportunities}, the opportunities and relevant problems of molecular communication is discussed. In~\cite{Wireless_Molecular}, several challenges that differentiate molecular communication with conventional wireless communication have been studied. Inter-symbol interference and channel coding are among the discussed issues. We feel that more fundamental studies are still required to understand the fundamental limits of molecular communication. As in~\cite{ISIT2011_Arash,ITW2011_Arash}, in this paper, we will consider encoding information in the concentration of molecules. However, we depart from~\cite{ISIT2011_Arash,ITW2011_Arash} in that a node is consisted of several bio entities for the reasons we discuss next. Further, we will use a different model for molecular reception.

The communication between two bio entities (e.g. bacteria) entails huge amount of randomness and, hence, is highly unreliable. This claim will become clear later through our result. Further, due to relying on chemical reactions to convey the information, the delay in the communication can be fairly large. Hence, one fundamental challenge is  how to form reliable communication between two unreliable bio entities (e.g. bacteria). To address this issue, rather than having molecular communication between two individual bio entities, we propose an architecture in which a cluster of biological entities (i.e. a cluster of bacteria) communicates with another. We will refer to this cluster of biological entities trapped in a chamber as a \emph{node}. The basic building blocks of the communication system are these clusters of bio entities which are able to transfer information from one point to another. Throughout this work, these bio entities are considered to be genetically modified bacteria~\cite{Danino2010,Palacios2011} which can sense specific types of molecules and respond accordingly. Although the principles of molecular communication is expected to hold for the bio entities, we particularly choose bacteria in our model as in~\cite{ISIT2012_Arash,ITW2012_Arash}. In short, an individual bacterium is very primitive and unreliable and, hence, incapable of providing reliable information exchange by itself. However, as we will study in details and quantify through information-theoretic tools, when a cluster of these bacteria form a biological node, they are collectively capable of reliable transmission and reception of information.

Molecular communication between two bio nodes consists of three processes: 1) molecule production 2) diffusion of molecules in the medium, and 3) molecule reception and GFP output production process.
The  concentration of produced molecules represents the information of the transmitter node. These molecules travel in the channel via diffusion process. They would be then received and decoded by the receiver node. The output of the receiver node, in the form of luminescence or fluorescence, is measured in steady-state to estimate the concentration of molecules at the vicinity of the node, and hence decode the transmitted information.

The goal is to compute the maximum amount of information that can be conveyed reliably per channel use. To this end,
\begin{itemize}
\item We model the input-output relation of each bacterium and use a probabilistic model to account for the  discrepancy in the behavior of individual bacterium within the population residing in a node.  
\item We obtain the optimal distribution on the input concentration (i.e., the molecular signal concentration) that results in maximum mutual information between the input and output (i.e., the maximum capacity).
\item Finally, we present an M-ary molecular signaling technique and obtain the resulting information rate and the corresponding error rate versus the maximum molecule (concentration) production. 
\end{itemize}

The rest of the paper is organized as follows. In Sec.~\ref{sec:model}, we present the problem and the model that we use throughout the paper. In Sec.~\ref{sec:transmitter}, the process of production of molecules at the transmitter is discussed. Sec.~\ref{sec:channel} describes briefly the diffusion of molecules through the channel followed by Sec.~\ref{sec:receiver} where we study the receiver functionality. The analysis of the capacity is discussed in Sec.~\ref{sec:capacity}. Then, Sec.~\ref{sec:modulation} introduces a practical molecular signaling scheme for the communication and analyzes its rate and reliability. Finally, Sec.~\ref{sec:conclusion} concludes the paper.

\section{Problem Statement and Bacteria Functionality Model}
\label{sec:model}

\begin{figure*}
\begin{center}
  \subfigure[A bacteria surrounded by AHL molecules]{
  \includegraphics[width=.4\textwidth]{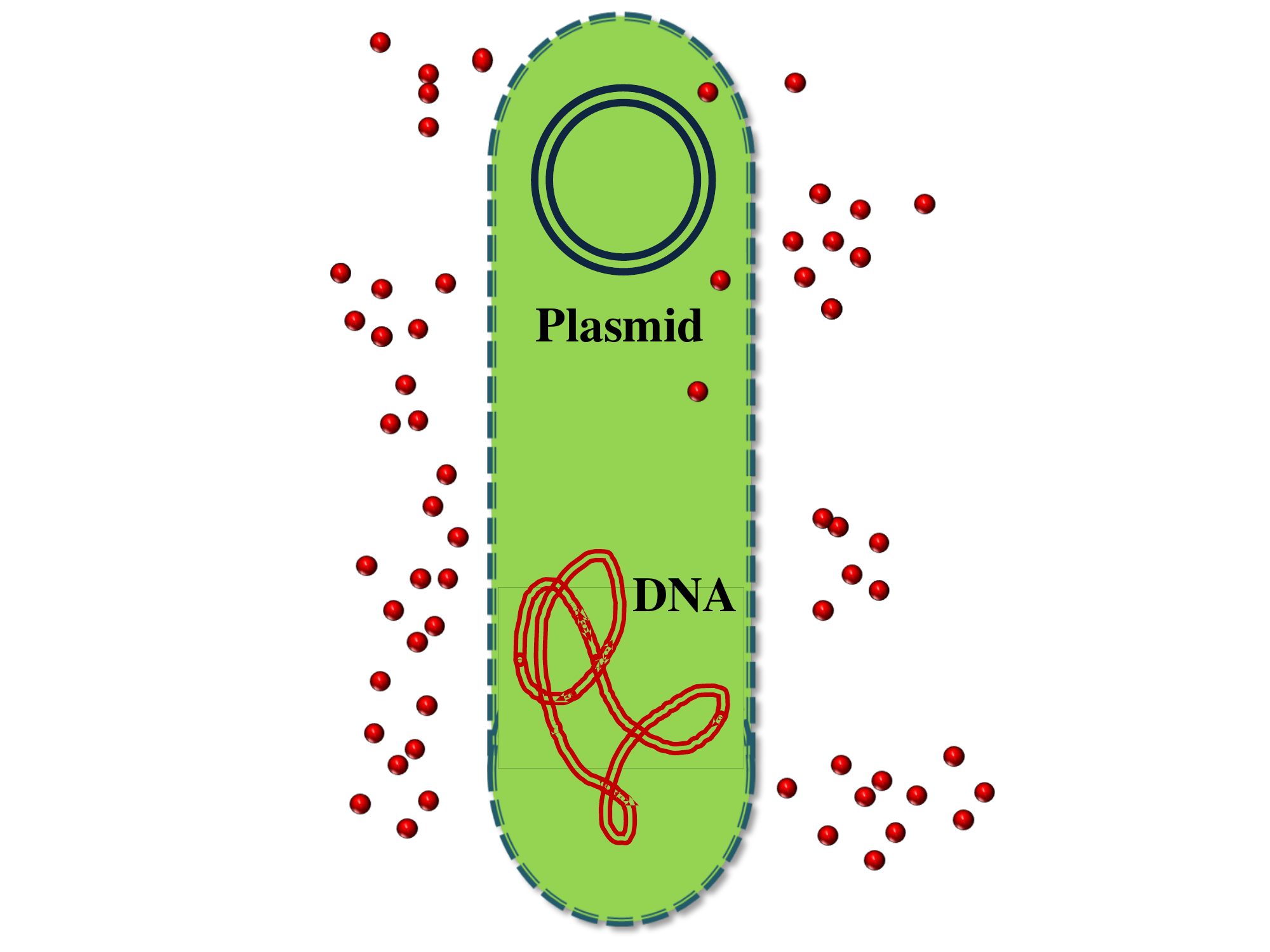}
  \label{fig:bacteria_model}
  }
  \subfigure[The main components in production of GFP by bacteria through reception of AHL molecules]{
  \includegraphics[width=.4\textwidth]{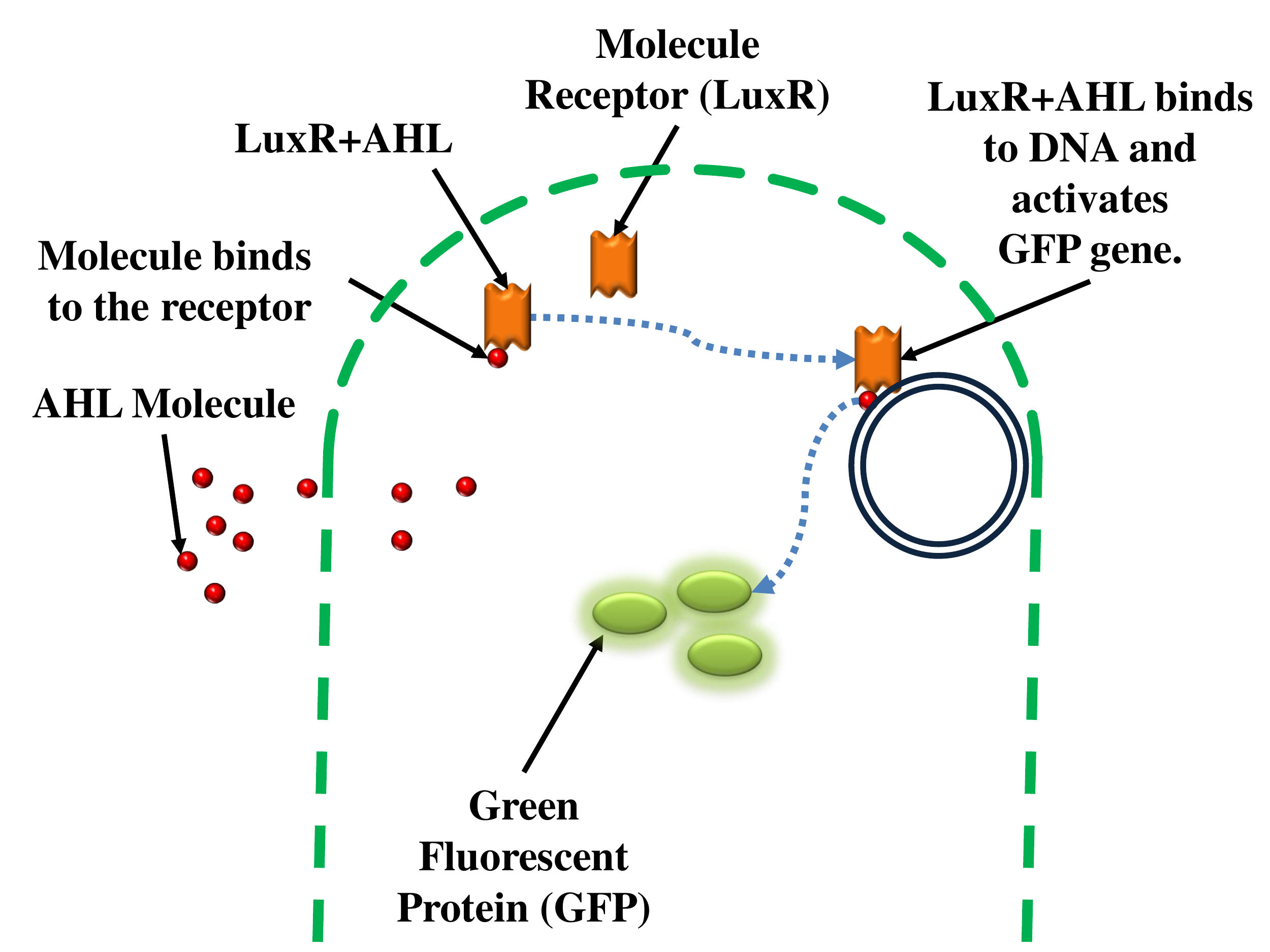}
  \label{fig:bacteria}
  }
\end{center}
\vspace{-.1in}
\caption{Schematic of a bacteria and processes involved in producing GFP}
\label{fig:network_model}
\end{figure*}

As explained in the introduction, we consider molecular communication between two nodes each containing $n$ bio entities (e.g. engineered bacteria$\footnote[1]{Hereafter, we refer to these bio-entities as bacteria, although the principles developed in this work is not confined to the bacteria.}$).  The information is conveyed by the transmitter node to the receiver node through the alternation of the concentration of Acyl Homoserine-Lactone (AHL) molecules surrounding the receiver node. This concentration is sensed by the bacteria in the receiver node; triggering each bacterium to possibly emit GFP with the intensity that depends on the concentration of AHL molecules. Fig.~\ref{fig:bacteria_model} shows a bacterium used in such a node. The features of the production of GFP (i.e., the proteins responsible for detection or production of molecules) is stored in the plasmid which can be added to a bacterium who does not naturally emit GFP. 

\begin{figure*}
\vspace{-.7in}
\centering
\includegraphics[width= .7\textwidth]{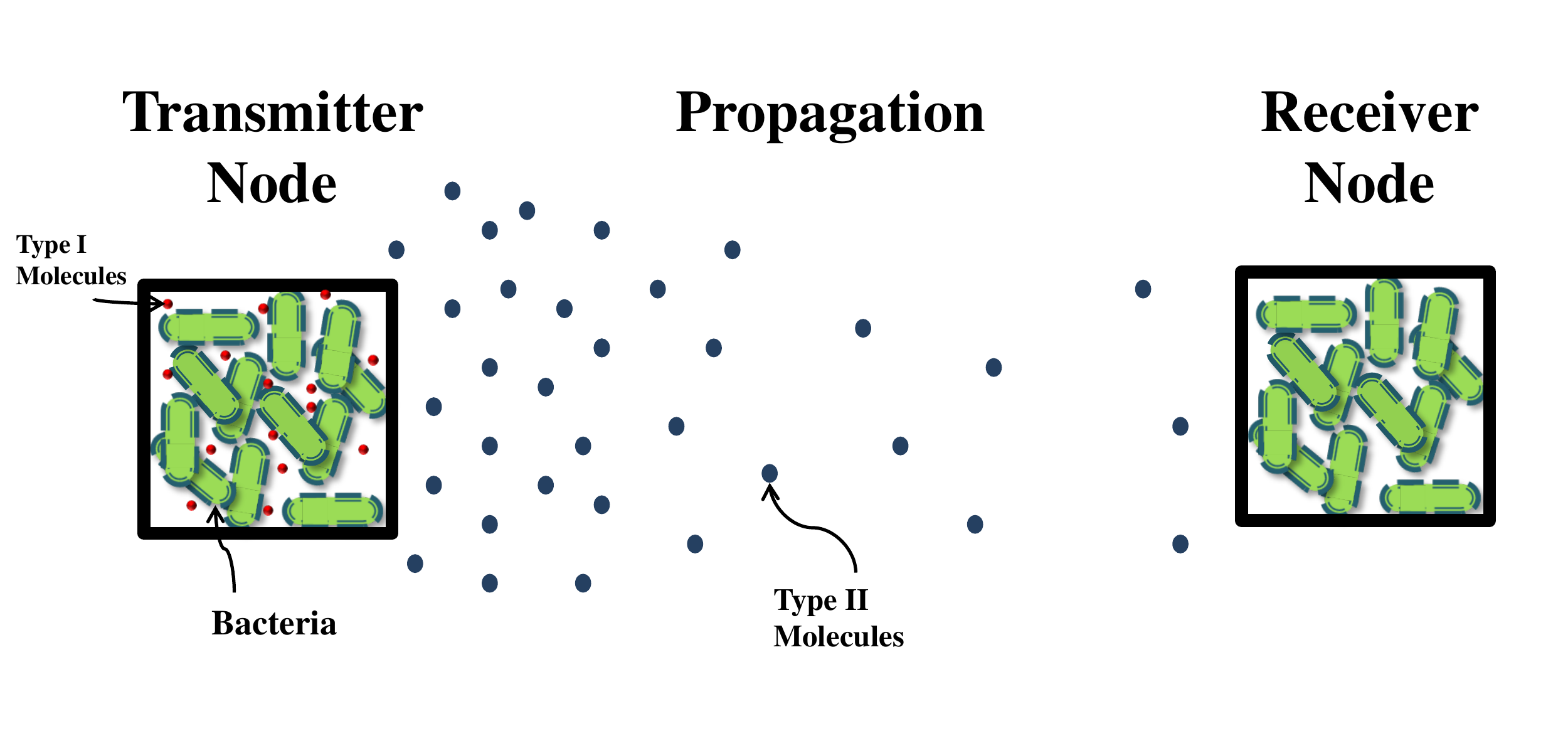}
\vspace{-.3in}
\caption{The molecular communication setup consisting of the transmitter, channel and the receiver}
\vspace{-.1in}
\label{fig:two_node}
\end{figure*}

The schematic for the communication between two nodes is depicted in Fig.~\ref{fig:two_node}.  Each node contains $n$ bacteria. Note that throughout the paper, we assume that the number of bacteria inside a chamber remains constant. The bacteria inside the chambers grow and divide. To maintain the population size constant, the new bacteria can be washed away from the chambers as in~\cite{Monaco}. Each bacterium is assumed to be able to sense and produce two different types of AHL molecules, namely type I and II \footnote[2]{The labeling of the molecular types is chosen for the convenience of the presentation. Please refer to~\cite{Bassler_Harveyi} for the technical labeling of these molecular signals.}. Let assume the transmitter node has some information in the form of concentration $A_0$ that would like to convey to the receiver node. The bacteria inside the transmitter node can be stimulated through its chamber with different levels of concentration of type I molecules in order to produce various concentrations of type II molecules to be propagated through the channel. The transmitter output as shown in the following has a probabilistic nature such that higher levels of molecule stimulant would, on average, result in higher levels of the output. The probabilistic nature of the molecule production introduces the first component of the uncertainty in the communication. The specifics of the transmitter node will be discussed in the next section. The emitted molecules would then diffuse through the channel to the receiver at the distance $r$ from the transmitter. The induced molecular concentration at the receiver node depends on the distance $r$. We assume that the transmitter node has an estimate $r_0$ for $r$ which maybe slightly different. That introduces the second component of the uncertainty in the communication from the transmitter to the receiver which will be explained in section~\ref{sec:channel}. At the receiver, as explained in section~\ref{sec:receiver}, each bacterium senses the concentration of type II molecules through the corresponding type II receptors. The reception of molecules again has a probabilistic nature which introduces the third component of the communication uncertainty. Upon the reception of molecules, a chain of reactions is triggered resulting in the production of GFP by bacteria.  The GFP output of the receiver node is used to decode the signal (i.e., the concentration $A_0$).

We intend to obtain the maximum information that the GFP output of the receiver node can give about the channel input concentration produced by the transmitter node. As such, we consider the models for the transmitter, the channel, and the receiver and analyze the uncertainty in each component which is aggregated with the noise from previous components. The  capacity of the communication will be obtained by considering all these three components together.

As studied earlier, to perform its functionality as a transmitter or a receiver, each bacterium must be able to sense either type I or type II molecules, respectively. Each bacterium might be equipped in general with two distinct receptor types: one for each molecule types. However, depending on the functionality, as a transmitter or receptor, only one type of receptors is enabled. We assume $N$ ligand receptors for each type of molecules. The bacteria response to these two stimulants (i.e., molecular types) is different, i.e., producing type II molecules upon the reception of type I but producing GFP in response to type II molecules. The functionality model for the reception of these two  molecule types is assumed to be the same, i.e., the process of reception is governed by the same set of equations with possibly different coefficients.

Fortunately, the response of various strands of bacteria to different levels of inter-cellular AHL molecule concentrations has been already studied extensively. We will adopt those mathematical models accounting for the chain of chemical reactions inside the bacterium due to the AHL molecular stimulus. 
In~\cite{muller2008}, a model consisting of a chain of linear differential equations is introduced to account for the production of luminescence or fluorescence in response to the presence of AHL molecules in the medium. These equations capture the average dynamic behavior of bacteria and also their steady-state behavior. They account for three main phases in the process as shown in Fig.~\ref{fig:bacteria}:
\begin{enumerate}
\item Binding of AHL molecules to the cell receptors (i.e., the ligand receptors)
\item Production of the AHL+LuxR complex and transcription of genes responsible for the generation of GFP, and finally, 
\item GFP production. 
\end{enumerate}
The differential  equation accounting for probability $p$ of binding of molecules to the cell receptors is given by~\cite{muller2008}:

\begin{equation}
\label{eq:diff}
\dot{p}=-\kappa p+A\gamma \left(1-p\right),
\end{equation}
where $A$ is the concentration of molecules surrounding the bacterium, $\gamma$ is the input gain and $\kappa$ is the dissociation rate of trapped molecules in the cell receptors. Here, $\dot{p}$ is the derivative of $p$ with respect to time. In this model, each cell receptor is activated (via a trapped molecule) with a probability that depends on the concentration of molecules in the medium surrounding the cell. According to~(\ref{eq:diff}), this probability starts growing from the moment a constant concentration $A$ is applied until it takes its final steady-state value $p^*$, given by
\begin{equation}
\label{eq:steady_state}
p^*=\frac{A\gamma}{A\gamma+\kappa} 
.\end{equation}
 Note that $p^*$ increases monotonically with respect to $A$ and approaches to $1$ for very high concentrations.

 The production of complex molecules, the transcription of genes and the production of GFP are modeled similarly as~\cite{muller2008}:
\vspace{-.05in}
\begin{equation}
\left\{
\begin{array}{rcl}
\dot{S}_1 &=& (b_0p+a_0)-b_1 S_1\\
\dot{S}_2 &=& a_1 S_1-b_2 S_2
\end{array}
\right.
\label{eq:chain_diff2}
.\end{equation}
where $S_1$ and $S_2$ are two post-transcription messengers and $b_i$ and $a_i$ are some constants~\cite{muller2008}. 
In the steady state, the final product $S_2$ is obtained as
\begin{equation}
\label{eq:s2_steady}
S_2^*=\frac{a_1(b_0p^*+a_0)}{b_1 b_2}
\end{equation}

Note that $S_2^*$, i.e., the intensity of GFP, has an initial value even when $p^*=0$ and increases linearly when $p^*$ increases. In the following, we use the above results to model the behavior of different components of the molecular communication system.

\section{The Transmitter Model}
\label{sec:transmitter}

As discussed in the previous section, to generate the desired type II concentration $A_0$ at the receiver, the bacteria residing in the transmitter node must be stimulated with type I molecules with the appropriate concentration $A_s$. We assume that the noise  in the transmitter output (i.e., the aggregated type II concentration $A_0$ at the receiver due to all $n$ bacteria residing in the transmitter node) is originated from the discrepancy in the individual behavior of the bacteria in the transmitter node. In other words, even though the average behavior of bacteria can be described by the set of deterministic differential equations in section~\ref{sec:model}, the individual behavior of bacteria features randomness. Such randomness can be accounted for by treating the constants in~(\ref{eq:diff}) through~(\ref{eq:s2_steady}) as random variables.

Two factors contribute to the uncertainty of the molecular concentration output of the transmitter node. One is the probabilistic nature of the number of activated receptors within a single bacterium in response to the type I stimulus. We model this by assuming each receptor being active as a Bernoulli random variable that is $1$ with probability $p^*$ defined in~(\ref{eq:steady_state}). The other factor is the randomness in $p^*$ itself from one bacterium to another within the node; due to the variation of the constant parameters (i.e., $\gamma$ and $\kappa$) in (\ref{eq:steady_state}) among the population of bacteria. Note that the other parameters, in addition to $\gamma$ and $\kappa$, in the model can be considered as random variables as well but the form of the final results in terms of communication capacity and reliability answer would remain the same. Hence, for simplicity of modeling the discrepancy of bacteria behavior in a node, we only consider the above parameters to contain randomness.
We model this variation with an iid additive noise $\epsilon_{\gamma}$ in the input gain $\gamma$ and $\epsilon_{\kappa}$ in the dissociation rate $\kappa$.  Hence, the entrapment probability $p_s$ upon the reception of the stimulant concentration $A_s$ by the bacteria at the transmitter would be given by
\begin{equation}
\label{eq:noisy_probability}
p_s=\frac{A_s(\gamma+\epsilon_\gamma)}{A_s(\gamma+\epsilon_{\gamma})+(\kappa+\epsilon_{\kappa})},
\end{equation}
where  $\epsilon_{\gamma}$ and $\epsilon_{\kappa}$ are zero-mean Normal noises with variances $\sigma_{\gamma}^2$ and $\sigma_{\kappa}^2$, respectively. The variance of these noises is assumed to be sufficiently small such that we can ignore the second and higher orders of $(\frac{\epsilon_{\gamma}}{\gamma})$ and $(\frac{\epsilon_{\kappa}}{\kappa})$ in the Taylor expansion of~(\ref{eq:noisy_probability}). We assume the same $p_s$ for all the receptors belonging to the same bacterium, but it varies from one bacterium to another in a node, as we described in~(\ref{eq:noisy_probability}).
 We define the noiseless entrapment probability at the transmitter as
\begin{equation}
\label{eq:noiseless_probability}
 p_s^*=\frac{A_s \gamma}{A_s \gamma+\kappa}.
\end{equation}
Hence, by approximating~(\ref{eq:noisy_probability}) as described above, we  have
\begin{equation}
\label{eq:noisy_probability2}
p_s \simeq p_s^*+p_s^*(1-p_s^*) \frac{\epsilon_{\gamma}}{\gamma}-p_s^*(1-p_s^*) \frac{\epsilon_{\kappa}}{\kappa} .
\end{equation}
The total number of activated receptors of $i^{th}$ bacterium, $X_i$, is a Binomial random variable with parameters $(N,p_{s,i})$ where $p_{s,i}$ is the realization of $p_s$ for the $i^{th}$ bacterium. Recall that $N$ is the number of ligand receptors per bacterium for a given molecule type. We denote $X$ as the total number of activated receptors of all the bacteria in the transmitter node. Hence, $X=\sum_{i=1}^n X_i $. 
 Using the conditional expectation, we have
\begin{equation}
\label{eq:expected_value}
E(X_i)=E(E(X_i|p_{s,i}))=E(Np_{s,i})=Np_s^*, \nonumber
\end{equation}
where the last equality is due to the fact that the $\epsilon_{\gamma}$ and $\epsilon_{\kappa}$ have zero means. Hence, we have
\begin{equation}
\label{eq:expected_output}
 E(X)=n N p_s^*=nN\frac{A_s \gamma}{A_s \gamma+\kappa}.  
\end{equation}
By using the conditional variance, we have
%
\begin{eqnarray}
\label{eq:variance}
Var(X_i) &=& E(Var(X_i|p_{s,i}))+Var(E(X_i|p_{s,i})) \nonumber\\
	\nonumber&=& E(N p_{s,i}(1-p_{s,i}))+Var(Np_{s,i})\\
&=& N p_s^*(1 -p_s^*) \nonumber \\ &&+(N^2-N){p_s^*}^2 (1-p_s^*)^2 (\frac{ \sigma_{\gamma}^2}{\gamma^2}+\frac{ \sigma_{\kappa}^2}{\kappa^2}).
\end{eqnarray}

The first term in~(\ref{eq:variance}) is due to the general uncertainty in a Binomial random variable (i.e., the probabilistic nature of the ligand reception) and the second term is due to the noise in the parameter $p_s$. Assuming independency among the behavior of different bacteria, the variance of the total number of activated receptors at the transmitter node is obtained as
\begin{eqnarray}
\label{eq:variance_output}
 Var(X)&=&n N p_s^*(1 -p_s^*)  \\ &&+ n(N^2-N) {p_s^*}^2 (1-p_s^*)^2 (\frac{ \sigma_{\gamma}^2}{\gamma^2}+\frac{ \sigma_{\kappa}^2}{\kappa^2}) \nonumber. 
\end{eqnarray}
Since the number of receptors $N$ per bacterium is usually large enough, the second term is dominating. Hence, we can approximate the variance by
\begin{equation}
\label{eq:variance_approximation}
 Var(X)\simeq n N^2  {p_s^*}^2 (1-p_s^*)^2 (\frac{ \sigma_{\gamma}^2}{\gamma^2}+\frac{ \sigma_{\kappa}^2}{\kappa^2}). 
\end{equation}

As observed in~(\ref{eq:s2_steady}), the average output of bacteria has an initial value corresponding to $p^*=0$ and increases linearly with the average number of activated receptors. We assume that this offset value is the same for all the bacteria in a population and independent of the input concentration. Hence, the measured output is considered to be the GFP production due to the presence of AHL molecules, which depends linearly on the number of activated receptors $X$.
 Therefor, the total type II molecular output  of the transmitter node is equal to $\alpha X$ where $\alpha$ is associated with a single activated receptor. Using~(\ref{eq:s2_steady}), we note that $\alpha N=\frac{a_1b_0}{b_1b_2}$. To the rest of paper, we consider $\alpha X$ as the output of the transmitter node.

In order to make the analysis tractable, we use Central Limit Theorem (CLT) to approximate $ X$, which is a Binomial random variable with known mean and variance, as a Normal random variable. In other words, since the number of receptors $N$ is large, we can use CLT to approximate $X_i$ by ${\bf \mathcal{N}} (Np_s^*,Var(X_i))$ where $Var(X_i)$ is given in~(\ref{eq:variance}). Hence, $X$ would  be the sum of $n$ Normal variables given by
\begin{equation}
\label{eq:normal_output} 
X=nNp_s^*+\epsilon_X,
\end{equation}
where $p_s^*$ is given in~(\ref{eq:noiseless_probability}) and $\epsilon_X$ has a Normal distribution $\mathcal{N}(0,Var(X))$ where $Var(X)$ is given by~(\ref{eq:variance_approximation}). The emitted molecules are propagated through the diffusion channel and reach the receiver . In the next section, we introduce the diffusion channel and study its effect in the communication.

\section{Diffusion Channel}
\label{sec:channel}

The molecules produced by the transmitter travel through the channel via the diffusion process. We characterize the temporal and spatial variations of molecules in the channel by following the general diffusion equations. According to Fick's second law of diffusion the concentration of molecules $A(r,t)$ at position $r$ at time $t$ is computed as follows~\cite{random_walk}:
\begin{equation}
\label{eq:fick_partial}
 \frac {\partial A(r,t)}{\partial t} = D \nabla^2 A(r,t).
\end{equation}
Here, $r$ is the distance of any point in the environment from the source, $D$ is the diffusion coefficient of the medium, and $\nabla$ is the del operator. The impulse response of~(\ref{eq:fick_partial}), is the Green's function $g_d(r,t)$ whose expression is as follows:
\begin{equation}
\label{eq:green_function}
 g_d(r,t)=\frac {1}{(4\pi Dt)^{\frac{3}{2}}} \exp{\left(-\frac{r^2}{4Dt}\right)}.\
\end{equation}
This impulse response is given for the 3-D medium using the observation that 3-D diffusion is equivalent to $3$ separate (simultaneous) 1-D diffusions. Since the diffusion equation is a linear equation, the solution to~(\ref{eq:fick_partial}) for an arbitrary channel input concentration $F(r,t)$ can be obtained using  
\begin{equation}
\label{eq:diffusion_convolution}
 A(r,t)=g_d(r,t)\otimes F(r,t),
\end{equation}
where $\otimes $ denotes a multi-dimensional convolution operation on $r$ and $t$.

In our setup, we assume that there is only one transmitter node emitting molecules. Therefore, in~(\ref{eq:diffusion_convolution}), we have $F(r,t)=f(t)\delta(r)$, where $f(t)$ is the time-dependent channel input signal. Further, assume that the channel is stimulated with a constant molecule rate $\beta$ for the duration $t_0$. In other words, the channel input rate is a constant concentration of $\beta$ molecules per unit of volume for $0\leq t \leq t_0$. Hence, we have 
\begin{eqnarray}
\label{eq:diffusion_response}
A(r,t) &=& \int^{t}_0 \beta\frac {1}{(4\pi D\tau)^{\frac{3}{2}}} \exp{\left( -\frac{r^2}{4D\tau}\right)} \, \mathrm{d}{\tau} \nonumber\\
&=& \frac{\beta}{4\pi D r} \text{erfc}\frac{r}{(4Dt)^{\frac{1}{2}}} \;\;\;\; 0\leq t<t_0
\end{eqnarray}
where erfc($x$) is the error function complement( i.e.,1-erf($x$)). Note that the error function erf($x$) is defined by the integral 
\begin{equation}
\label{eq:error_function}
erf(x)=\frac{2}{\sqrt{\pi}} \int_0^x e^{-u^2} du.
\end{equation}

Note that  for $t_0<t$ the concentration of molecules can be obtained from
\begin{equation}
\label{eq:after_response}
A(r,t)= \frac{\beta}{4\pi D r}( \text{erfc}(\frac{r}{(4Dt)^{\frac{1}{2}}})-\text{erfc}(\frac{r}{(4D(t-t_0))^{\frac{1}{2}}}))
\end{equation} 
Since $ \text{erfc}\frac{r}{(4Dt)^{\frac{3}{2}}}$ approached $1$ for large values of $t$, the pulse duration $t_0$ must be long enough to allow the channel output concentration in~(\ref{eq:diffusion_response}) to approach its steady-state value given by
\begin{equation}
\label{eq:concentration_steady_state}
A^*(r)=\frac{\beta}{4\pi D r}
\end{equation}
This response is valid for open free medium in which the only boundary conditions are at the transmitter. If the dimension $R$ of the receiver node is comparable to the distance $r$ between the nodes, a factor $(1-\frac{R}{r})$ should be multiplied to the expression in~(\ref{eq:concentration_steady_state}). In this paper, we assume that the distance between the transmitter and receiver nodes is significantly larger than the size of the nodes. Hence, we ignore the effect of this term on the steady-state response.

As discussed in the previous section, each activated receptor contributes an amount of $\alpha$ to the output rate.  Since the diffusion channel is linear, the total response at the receiver will be the superposition of individual responses. Hence, $\beta=\alpha X$ and the steady-state concentration $A_r$ at the receiver is given by
\begin{equation}
\label{eq:concentration_R}
A_r=\frac{\alpha X}{4\pi D r},
\end{equation}
where $r$ is the distance between the transmitter and the receiver nodes. The response in (\ref{eq:concentration_R}) is the average response of Brownian motion of the individual molecules without any other interferences~\cite{max_diffusion}. Here, we consider another source of uncertainty which potentially arises in molecular communication due to uncertainty on the distance $r$ of the two nodes. We denote by $r_0$ the transmitter estimate for $r$, which maybe slightly different from $r$. Hence, we have
\begin{equation}
\label{eq:channel_noise}
A_r=\frac{\alpha X}{4\pi D (r_0+\epsilon_r)}\simeq \frac{\alpha X}{4\pi D r_0} (1-\frac{\epsilon_r}{r_0}),
\end{equation}
where $\epsilon_r$ is a zero-mean Normal random variable with variance $\sigma_r^2$ which is assumed to be much smaller than $r_0^2$. Therefore, we only considered the first order term $\frac{\epsilon_r}{r_0}$ in~(\ref{eq:channel_noise}). By using~(\ref{eq:expected_output}), we obtain
\begin{equation}
\label{eq:expected_input_receiver}
 E(A_r)=\frac{\alpha nN}{4\pi D r_0} \frac{A_s \gamma }{A_s \gamma+\kappa}.
\end{equation}
Hence, the required stimulating concentration $A_s$ for type I molecules at the transmitter can be obtained by equating the expected concentration of molecules at the receiver~(\ref{eq:expected_input_receiver}) to $A_0$. As described in the previous section, $A_0$ is the desired concentration of type II molecules to be transferred from the transmitter to the receiver node. Hence, we have
\begin{equation}
A_s=\frac{\kappa A_0}{\gamma (\frac{\alpha nN}{4\pi D r_0}-A_0)}.\nonumber
\end{equation}

From~(\ref{eq:normal_output}) and using the fact that $A_0=\frac{\alpha nN p_s^*}{4\pi Dr_0}$, we can write~(\ref{eq:channel_noise}) as
 \begin{align}
\label{eq:noisy_concentration}
A_r &=(A_0+ \frac{\alpha}{4\pi D r_0}\epsilon_X)(1-\frac{\epsilon_r}{r_0})\nonumber\\
&\simeq A_0+A_0 \epsilon_t-\frac{A_0}{r_0}\epsilon_r,
\end{align}
where we have ignored the second-order noise terms and $\epsilon_t$ is a zero-mean Normal random variable with variance given by
\begin{equation}
\label{eq:variance_t} 
\sigma_t^2= \frac{ (1-p_s^*)^2}{n} (\frac{\sigma_{\gamma}^2}{\gamma^2}+\frac{\sigma_{\kappa}^2}{\kappa^2}).
\end{equation}

The first term in~(\ref{eq:noisy_concentration}) can be viewed as the signal to be decoded by the receiver node. The second and third terms are signal dependent additive Gaussian noises due to the randomness at the transmitter and the channel uncertainty, respectively. We refer to these two noises as the transmitter and the channel noise perceived at the receiver in the molecular communication, respectively.

\section{Receiver Model}
\label{sec:receiver}

The concentration $A_r$ derived in~(\ref{eq:noisy_concentration}) is sensed by the bacteria in the receiver node. The sensing process of type II molecules is similar to that of the type I molecules we analyzed for the transmitter, but through different receptors. Hence, it follows the same chain of differential equations as in Section~\ref{sec:transmitter}. The difference is that the input concentration is noisy itself which introduces an additional uncertainty to the output of the receiver node; which is in the form of  GFP. This output is used to decode the information sent by the transmitter. Note that we assumed an ideal GFP detection system. However, in practice, the GFP detection system may introduce an additional uncertainty to the overall communication system. One may refer to\cite{GFP_Detection} where the GFP sensors are studied in details.

Here, we incorporate the effect of both noises introduced in the previous section in addition to the uncertainty contributed by the reception process itself. Again, assume the noises $\epsilon_{\gamma}$ and $\epsilon_{\kappa}$ with variances $\sigma_{\gamma}^2$ and $\sigma_{\kappa}^2$ account for the dependencies of gain $\gamma$ and the parameter $\kappa$ among the bacteria at the receiver node, respectively. In other words, the variations in the behavior of bacteria at the receiver node are incorporated by $\epsilon_{\gamma}$ and $\epsilon_{\kappa}$. Hence, the entrapment probability of type II molecules by a receptor at the receiver can be written as
\begin{equation}
\label{eq:noisy_probability1}
p_r=\frac{( A_0+A_0\epsilon_t-\frac{A_0}{r_0}\epsilon_r)(\gamma+\epsilon_{\gamma})}{(A_0+A_0\epsilon_t-\frac{A_0}{r_0}\epsilon_r)(\gamma+\epsilon_{\gamma})+(\kappa+\epsilon_{\kappa})}.
\end{equation}
Note that the input concentration noises $\epsilon_t$ and $\epsilon_r$ affect all the receiver bacteria in the same manner but $\epsilon_{\gamma}$ and $\epsilon_{\kappa}$ which account for the reception process of bacteria, are independent for different bacteria.
By approximating~(\ref{eq:noisy_probability1}) and again keeping only the first-order terms of the noises, we obtain
\begin{equation}
\label{eq:noisy_probability2}
p_r\simeq p_0+ p_0(1-p_0)( \frac{\epsilon_{\gamma}}{\gamma}  -\frac{\epsilon_{\kappa}}{\kappa}-\frac{\epsilon_r}{r_0}+ \epsilon_t),
\end{equation}
where we defined $p_0\triangleq \frac{A_0\gamma}{A_0\gamma+\kappa}$. The first term in the right hand side of~(\ref{eq:noisy_probability2}) is the ideal channel input. The noise terms in~(\ref{eq:noisy_probability2}) capture the uncertainty in all three components of communication, i.e., the molecule production at the transmitter, the diffusion in the channel, and the reception of molecules at the receiver. Note that $\epsilon_{\gamma}$, $\epsilon_{\kappa}$ and $\epsilon_r$ have constant variances $\sigma_{\gamma}^2$, $\sigma_{\kappa}^2$ and $\sigma_r^2$, respectively but the variance of $\epsilon_t$  given by~(\ref{eq:variance_t}) is signal dependent. Since the number of bacteria $n$ in a node is large, the variance in~(\ref{eq:variance_t}) is negligible relative to the other noise terms in~(\ref{eq:noisy_probability2}). In other words, the noise of the transmitter is effectively filtered in the reception process as it was expected due to the low-pass nature of the quorum sensing process as discussed in~\cite{muller2008}.

We denote by $Y_i$  the number of activated receptors of the $i^{th}$ bacterium in the receiver node at steady state. Then, $Y=\sum_{i=1}^n Y_i$ would give the total number of activated receptors of all $n$ bacteria in the node. Note that $Y$ is the sum of binomial random variables with parameters $(N,P_{r,i})$. Here, $p_{r,i}$ is the realization of $p_r$ for the $i^{th}$ bacterium. With a discussion similar to the transmitter in Sec.~\ref{sec:transmitter}, the GFP output of the receiver depends linearly on $Y$. Here, we consider $Y$ itself to be the output of the receiver.
The expected value of $Y$ can be obtained as
\begin{equation}
\label{eq:expected_final}
E(Y)=Nnp_0.
\end{equation}

Computing the variance of the output will be more involved. Since $\epsilon_r$ is the same for all the bacteria of a node, $Y_i$'s are independent given the value of $\epsilon_r$. Hence,

\begin{eqnarray}
Var(\sum_{i=1}^n Y_i|\epsilon_r) &=& \sum_{i=1}^n Var(Y_i|\epsilon_r) \nonumber \\
\label{eq:variance_light}
&\simeq&nN^2 (\frac{\sigma_{\gamma}^2}{\gamma^2}+ \frac{\sigma_{\kappa}^2}{\kappa^2}) (p_0-p_0(1-p_0)\frac{\epsilon_r}{r_0})^2 \nonumber \\
&& (1-(p_0-p_0(1-p_0)\frac{\epsilon_r}{r_0}))^2,
\end{eqnarray}
where the last equality is resulted by using $p_0-p_0(1-p_0)\frac{\epsilon_r}{r_0}$ as $p_0$ in~(\ref{eq:variance}) and neglecting $N$ relative to $N^2$. By keeping only the terms with the first order of $\epsilon_r$ and the assumption that $E(\epsilon_r)=0$, we obtain
\begin{equation}
\label{eq:variance_conditional}
E(Var(Y|\epsilon_r))=nN^2 (\frac{\sigma_{\gamma}^2}{\gamma^2}+ \frac{\sigma_{\kappa}^2}{\kappa^2}) p_0^2(1-p_0)^2.
\end{equation}
On the other hand, we have
\begin{align}
\label{eq:expected_conditional}\nonumber
Var\left (E(\sum_{i=1}^n Y_i|\epsilon_r )\right)&=Var\left(\sum_{i=1}^n N(p_0-p_0(1-p_0)\frac{\epsilon_r}{r_0})\right)\\
&=nN^2  p_0^2(1-p_0)^2 \frac{\sigma_r^2}{r_0^2}.
\end{align}
%
%
\begin{equation}
\label{eq:variance_final}
Var(Y)=nN^2 p_0^2(1-p_0)^2 (\frac{\sigma_{\gamma}^2}{\gamma^2}+ \frac{\sigma_{\kappa}^2}{\kappa^2}+\frac{\sigma_r^2}{r_0^2}).
\end{equation}

\section{Capacity Analysis}
\label{sec:capacity}

The analysis of the noisy Binomial random variable $Y$ is cumbersome. Hence, with the same argument as for the transmitter, we approximate $Y$ with a Normal random variable with the expected value and variance given by~(\ref{eq:expected_final}) and~(\ref{eq:variance_final}), respectively. Hence, the output of the receiver node would be in the form
\begin{equation}
\label{eq:receiver_output}
Y=nNp_0+\epsilon_Y,
\end{equation}
where $\epsilon_Y$ is a zero- mean normally distributed random variable with the mean and signal-dependent variance given by~(\ref{eq:expected_final}) and~(\ref{eq:variance_final}). Note that, the first term in~(\ref{eq:receiver_output}) is the  signal and the second term is the additive noise due to both the channel and the reception process. As described in the previous section, the noise due to the transmitter is neglected in $\epsilon_Y$ relative to the other two noises.  In addition, if we assume that the uncertainty in the position of the receiver is negligible (i.e., $\frac{\sigma_r^2}{r_0}$ is small due to sufficiently large $r_0$), then $\epsilon_Y$ is only due to the randomness in the reception of the molecules by the receiver. Note that throughout the discussion, we assumed that the number of bacteria $n$ remains constant inside the chambers.  The effect of variations of the number of bacteria would result in an additional variance term in the final output. However, it can be shown that this term is negligible compared to the other terms in the variance.
%

%
%

\begin{figure}[t]
\vspace{-1in}

\hspace{-.5in}
 \includegraphics[width=.58\textwidth]{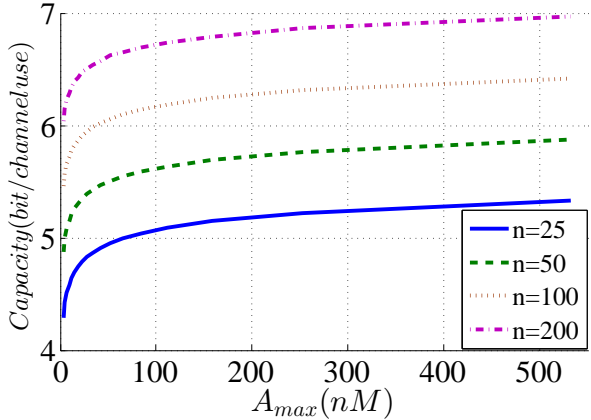} 
\vspace{-1.3in}
\caption{Capacity versus maximum concentration of molecules at the receiver $A_{max}$ for different numbers of bacteria in a node.}
\label{fig:capacity}
\end{figure}

\begin{figure}[t]
\vspace{-1in}
\hspace{-.5in}
\includegraphics[width=.58\textwidth]{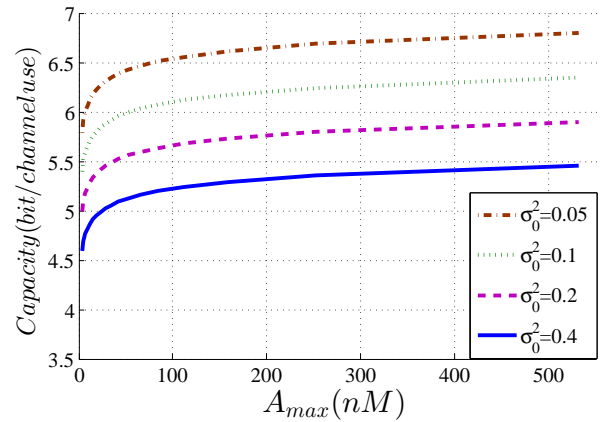}
\vspace{-1.3in}
\caption{Capacity versus maximum concentration of molecules at the receiver $A_{max}$ for different levels of noisel.}
  \label{fig:different_noise}
\end{figure}

In order to calculate the capacity per channel use from the transmitter to the receiver, we should obtain the optimized distribution of $A_0$ which maximizes $I(A_0;Y)$; the mutual information between the input and the output. Since $p_0$ is deterministically obtained from $A_0$ through~(\ref{eq:steady_state}), we can consider $p_0$ as the channel input and maximize the information that $Y$ gives about $p_0$:
\begin{equation}
\text{ max}_{f_{p_0}(p_0)} I(p_0;Y)=H(Y)-H(Y|p_0).
\end{equation}

The optimal distribution on $p_0$ would in turn give the optimized distribution of $A_0$. To proceed, we observe that, in practice, $A_0$ cannot take any arbitrary value due to the limitation in the molecule production of bacteria. Hence, we assume a maximum output concentration that is equal to $A_{max}$. This corresponds to probability $p_{max}=\frac{ A_{max}\gamma}{ A_{max} \gamma+\kappa}$ via~(\ref{eq:steady_state}). This maximum probability is due to the maximum concentration range with which the transmitter can emit the molecules into the medium. By using a higher range, the transmitter can increase the maximum concentration of molecules at the vicinity of the receiver node and increase $p_{max}$. Therefore, we obtain the optimized distribution for $p_0$ over the interval $ [0 \;\; p_{max}]$ and calculate the capacity with respect to $p_{max}$ or equivalently $A_{max}$.

The noise term in~(\ref{eq:receiver_output}) is complicated since the noise power depends on the signal itself as it can be seen in~(\ref{eq:variance_final}). Hence, we resort to use the numerical method of Blahut-Arimoto algorithm (BA) to obtain the optimal distribution for $p_0$ and its corresponding capacity.  Equation~(\ref{eq:variance_final}) implies that the noise power is at its maximum at $p_0=\frac{1}{2}$ and goes to zero when $p_0$ approaches to either zero or one. Hence, we expect that the distribution of $p_0$ should take values closer to $0$ and $p_{max}$ with a higher probability. The results from the numerical algorithm confirms this fact and the distribution has local maximums at $0$ and $p_{max}$.  
%

%
We define $\sigma_0^2\triangleq \frac{\sigma_{\gamma}^2}{\gamma^2}+ \frac{\sigma_{\kappa}^2}{\kappa^2}+\frac{\sigma_r^2}{r_0^2}$. Results for the capacity (in bits per sample) with respect to $A_{max}$, the maximum concentration of molecules which results in $p_{max}$, for different numbers of bacteria in the nodes is shown in Fig.~\ref{fig:capacity}. The unit of measurement used for the concentration of molecules is nano-Moles per litre (nM). In this setup, we assume $N=50$, $\sigma_0^2=0.1$ and also use the values $\kappa=0.1$, and $\gamma=0.0004$ from~\cite{muller2008}.
As we observe from the plot, the capacity increases when we increase $A_{max}$ which results in higher $p_{max}$  or increase the number of bacteria $n$. The ratio of the expected value of the output to its standard variance which is a measure of the decoding precision can be obtained using~(\ref{eq:expected_final}) and~(\ref{eq:variance_final}) as
\begin{equation}
\frac{E(Y)}{\sqrt{Var(Y)}}=\frac{\sqrt{n}}{(1-p_0) \sigma_0}.
\label{eq:ratio}
\end{equation}

As we observe in~(\ref{eq:ratio}), the ratio of the expected value and the standard deviation of the output increases as $\sqrt{n}$ and, hence, the capacity of the bacteria increases by using more bacteria in a node.  Note that the maximum achievable capacity is limited even if the transmitter used infinite $A_{max}$ to make $p_{max}=1$. 
The capacity for different values of $\sigma_0^2$ is shown in Fig.~\ref{fig:different_noise}. The results are shown for $n=100$ and $N=50$. 
 Note also that, in practice, $N$ and in particular $n$ can be very large. However due to the exponential growth of the time complexity of the numerical method with respect to $N$ and $n$, we only provided the capacity for relatively small values of $N$ and $n$.

\subsection{Information rate per time}

Thus far, we obtained the capacity per sample in the molecular communication system. In order to compute the information exchange rate per unit of time, we now study a Return-to-Zero communication scenario. In other words, the transmitter sends the information, waits until the information is received by the receiver, and finally the channel becomes empty and ready for the transmission of the next sample. In order to obtain the information rate per unit of time, we should consider both the delay imposed by the channel as well as the delay due to the the reception process. In particular, we obtain the time it takes for the concentration of molecules to reach the steady state at the receiver, the time it takes for the receiver bacteria to decode the concentration of molecules, and the time for the channel to become ready for transmission of the next sample. 

In order to account for the delay due to the channel, we need to obtain both the rise time $T_r$ and fall time $T_f$ of the diffusion process of the channel. The diffusion channel equations discussed in Sec IV (derived form Fick's second law) describe the average behavior of molecules (i.e., the concentration of molecules ) at each point at any time.
Note that the time for the channel to reach the steady-state response depends on the diffusion coefficient $D$ and the distance $r$. In particular, using~(\ref{eq:diffusion_response}) and~(\ref{eq:concentration_steady_state}), we obtain the ratio of the transient to the final steady-state response as 
\begin{equation}
\tau_r(r,t)=\frac{A(r,t)}{A^*(r)}=\text{erfc}\frac{r}{(4Dt)^{\frac{1}{2}}}
\label{eq:ratio}  
\end{equation}

\begin{figure}[t]
\vspace{-.5in}
\centering
\includegraphics[width=.45\textwidth]{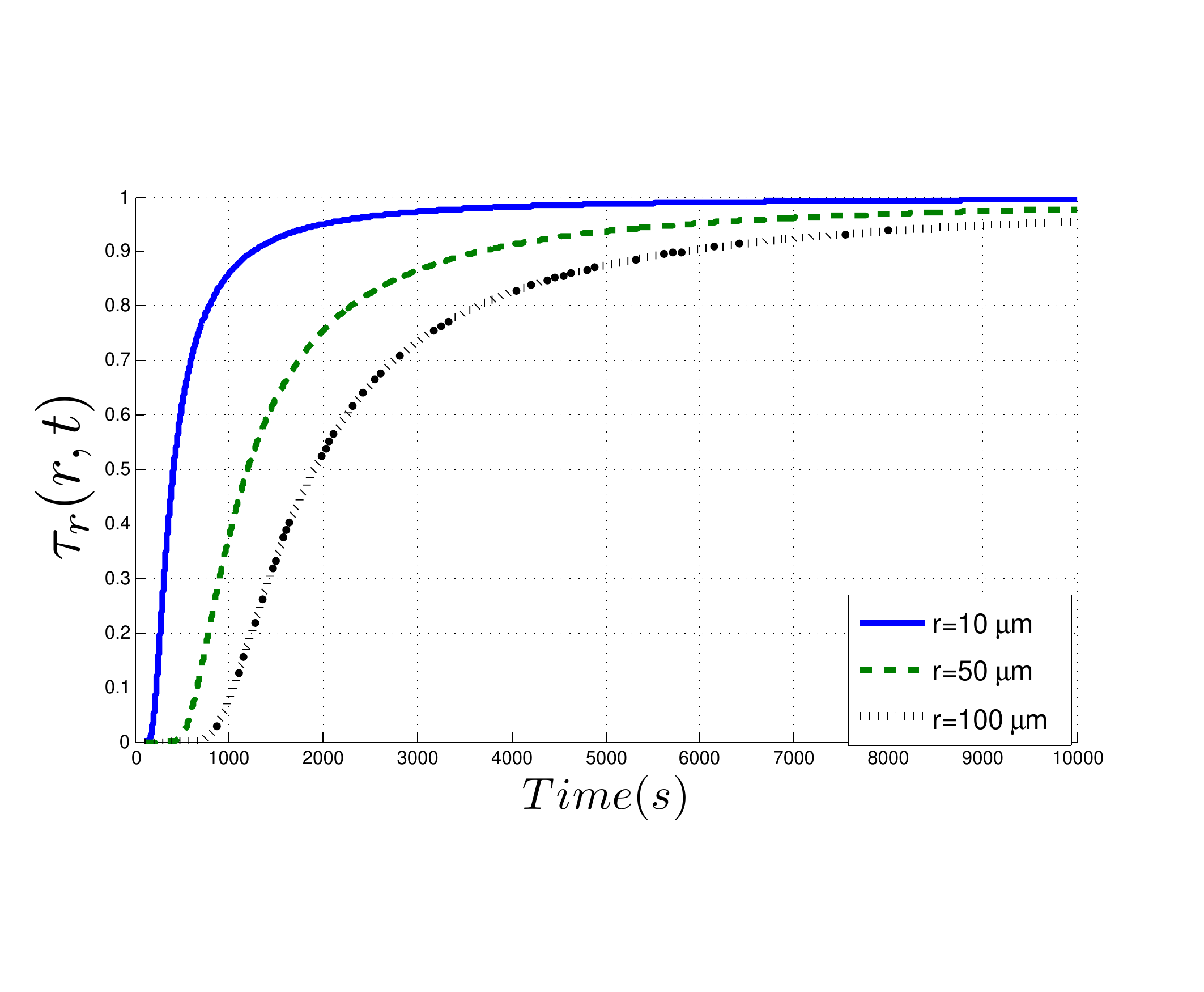}
\vspace{-.2in}
\caption{Rise time of the diffusion channel.}
\label{fig:rise}
\end{figure}

In Fig.~\ref{fig:rise}, we have plotted this ratio vs time for different distances between the transmitter and the receiver. In this plot, $D$ is assumed to be equal to $10^{-5}\; \frac{{cm}^2}{sec}$ which is a typical value for diffusion in water. The steady-state can be defined as the time that $\tau_r$ surpasses a specific threshold (e.g., $\tau_r=0.9$). 

On the other hand, in order to account for the fall time $T_f$ of the channel, we define by $\tau_f$ the ratio between the falling concentration of molecules and its maximum value at the steady state. Using~(\ref{eq:after_response}) and~(\ref{eq:concentration_steady_state})), we have:
\begin{equation}
\tau_f(r,t)=\frac{A(r,t)}{A^*(r)}=\text{erfc}(\frac{r}{(4Dt)^{\frac{1}{2}}})-\text{erfc}(\frac{r}{(4D(t-t_0))^{\frac{1}{2}}})
\label{eq:ratio}  
\end{equation}

\begin{figure}[t]
\vspace{-.5in}
\centering
\includegraphics[width=.45\textwidth]{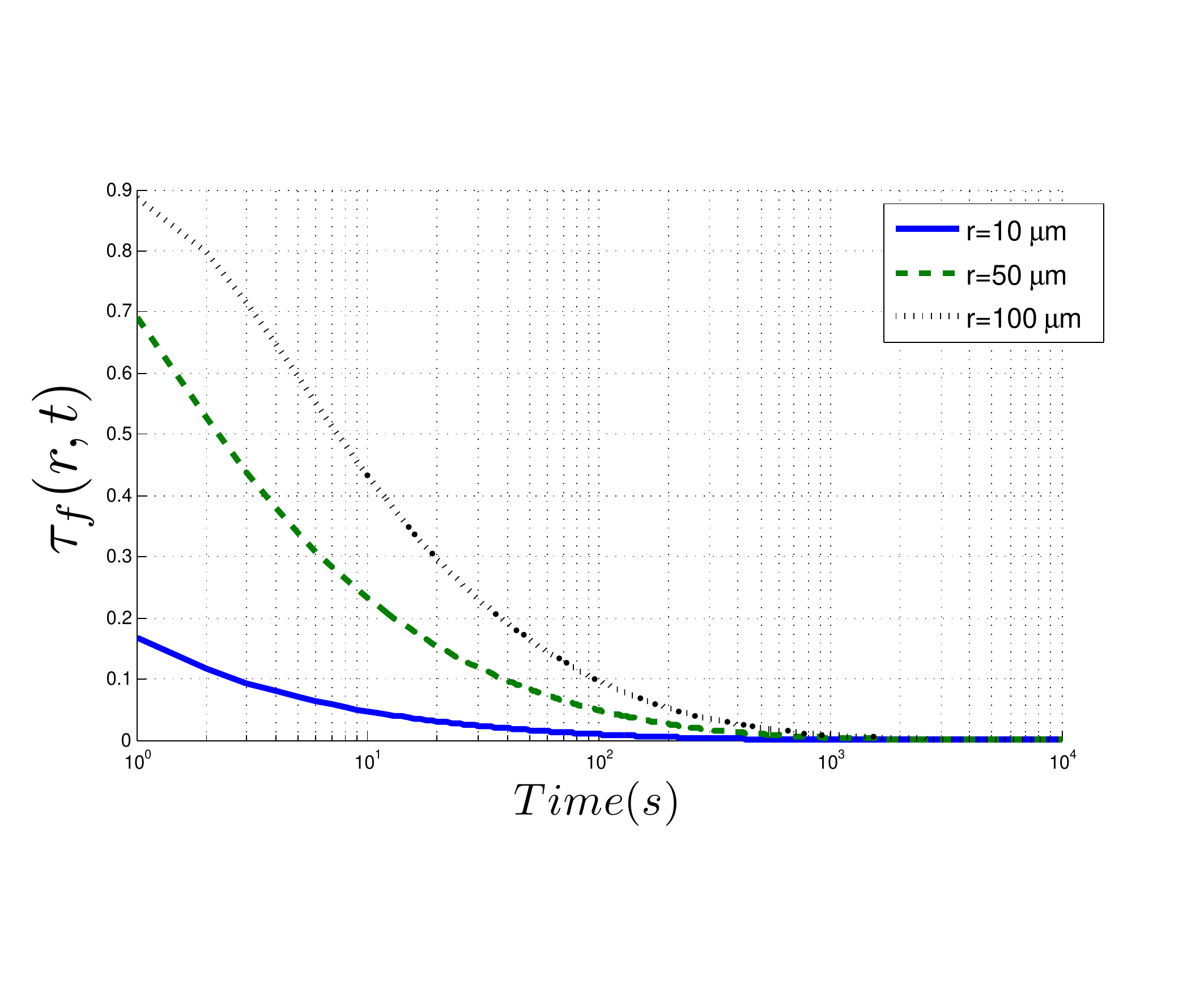}
\vspace{-.2in}
\caption{Fall time of the diffusion channel.}
\label{fig:fall}
\end{figure}

We assume the fall time of the channel to be the time that  $\tau_r$ falls below a specific threshold (e.g., $\tau_r=0.1$). We have plotted this ratio vs time for different distances in Fig.~\ref{fig:fall}. The aggregation of the rise time and fall time gives the total delay imposed by the channel in the communication from the transmitter to the receiver. As we see in the plots, both the rise time and the fall time are in the order of tens of minutes and  increases with the distance $r$.

The second component in the communication delay is the time it takes for the receiver bacteria to  receive and decode the concentration of molecules. Note that the concentration of molecules at the receiver should remain constant until decoded by the receiver.  Here, we use the set of differential equations introduced in the Sec. II. Moreover, we use the estimated parameters given in~\cite{muller2008} to obtain the value of the reception delay. As described in Sec. II, the total delay is due to the three processes that occur consecutively: entrapment of molecules, production of the proteins and transcription of genes, and  production of the GFP. Using (1), the time constant for the entrapment of molecules is obtained as $T_1=\frac{1}{A\gamma+\kappa}$. For a typical value of concentration of molecules (e.g., A=100nM), we have $T_1 \simeq 5 \; min$. The time constant for the production of protein and transcription of genes, obtained from the first equation in (3), is equal to $T_2=\frac{1}{b_1}\simeq 1\; hr$ . Finally, the time constant for production of fluorescent is obtained from the second equation in (3) and is equal to $T_3=\frac{1} {b_2}\simeq 10\; min $.  
 Due to the exponential behavior of the differential equations in (3), we consider three times of these time-constants as the time required to reach the steady-state response. Hence, the overall delay is approximated as $T_R\simeq 3\; hr$. Note that the production of proteins and transcription of genes are the dominating factors in the response delay and they are independent of the concentration of molecules $A$. Hence, we can assume that the above analysis holds true for all the values of $A$ in the range. 

Based on the discussion above, the total time it takes for the molecules to propagate through the channel and decoded by the receiver can be obtained by accumulating the delays discussed above. In other words, the total delay is given by $T_T=T_r+T_R+T_f$ and is in the order of a few hours. This is the time the transmitter should wait before sending the next sample. Hence, the information rate per unit of time can be obtain by dividing the rate per channel shown in Fig. 3 by the total delay $T_T$.  Moreover, note that the stimulation time $t_0$ introduced in Section~\ref{sec:channel} would be equal to $T_r+T_R$.
The information rate per hour is shown in Table 1 for different values of $n$ in Fig 3, and variations of $r$ in Fig.~5 and 6. We assumed $\sigma_0^2=0.1$ and $A_{max}=400 nM$.
\begin{table}
\begin{center}
    \begin{tabular}{| l | l | l | l |}
    \hline
    bits per hour& $r=10\mu$ &$r=50\mu$  & $r=100\mu$  \\ \hline
    $n=50$ & 1.7 & 1.4 & 1.2 \\ \hline
    $n=100$ & 1.9 & 1.6 & 1.3 \\ \hline
    $n=200$ & 2.1 & 1.8 & 1.4  \\
    \hline
    \end{tabular}
\vspace{.15in}
\caption{The information rate per hour}

\end{center}
\vspace{-.15in}
\end{table}

As we see in the table, low information rates (1-2 bits per hour) can be achieved in the diffusion-based molecular communication by trading the rate for reliability. One option to increase the rate is to use much larger number of bacteria. As we note from Fig.~3, the rate increases almost logarithmically in $n$.

\section{Achievable Rates under M-Ary Signaling (Modulation)}
\label{sec:modulation}

The analysis in the previous section was based on the assumption that any continuous values of the concentration less than $A_{max}$ can be produced and received by the nodes. By that assumption we obtained the maximum amount of information that can be communicated per channel use. In this section, we consider a specific signaling method (i.e., M-ary modulation) and study the information exchange rate and the corresponding achievable error rate. We use only a finite discrete number of levels of molecular concentrations $A_0$ for communication which results in discrete levels of $p_0$.
 The range of the input is determined by $p_{max}$. Two factors influence the signaling performance: the number of levels of molecular concentration and the choices for the values of those levels. 

We consider the scenario in which $m$ symbols to be chosen with uniform spacing from the interval $[0\quad p_{max}]$. Therefor, the $i^{th}$ symbol level would correspond to $p_{max} \frac{i}{m-1}, 0\leq i\leq m-1$. We show by $p_{e,i}$ the probability of error in the detection of $i^{th}$ symbol. Hence, the total probability of error is equal to $p_e=\sum_{i=0}^{m-1} w_i p_{e,i}$, where the weights $w_i$ associated with the $m$ symbols must be obtained.

\begin{figure*}
\centering
  \subfigure[The information rate versus the maximum concentration of molecules at the receiver $A_{max}$ for different M-ary schemes]{
 \includegraphics[width=.48\textwidth]{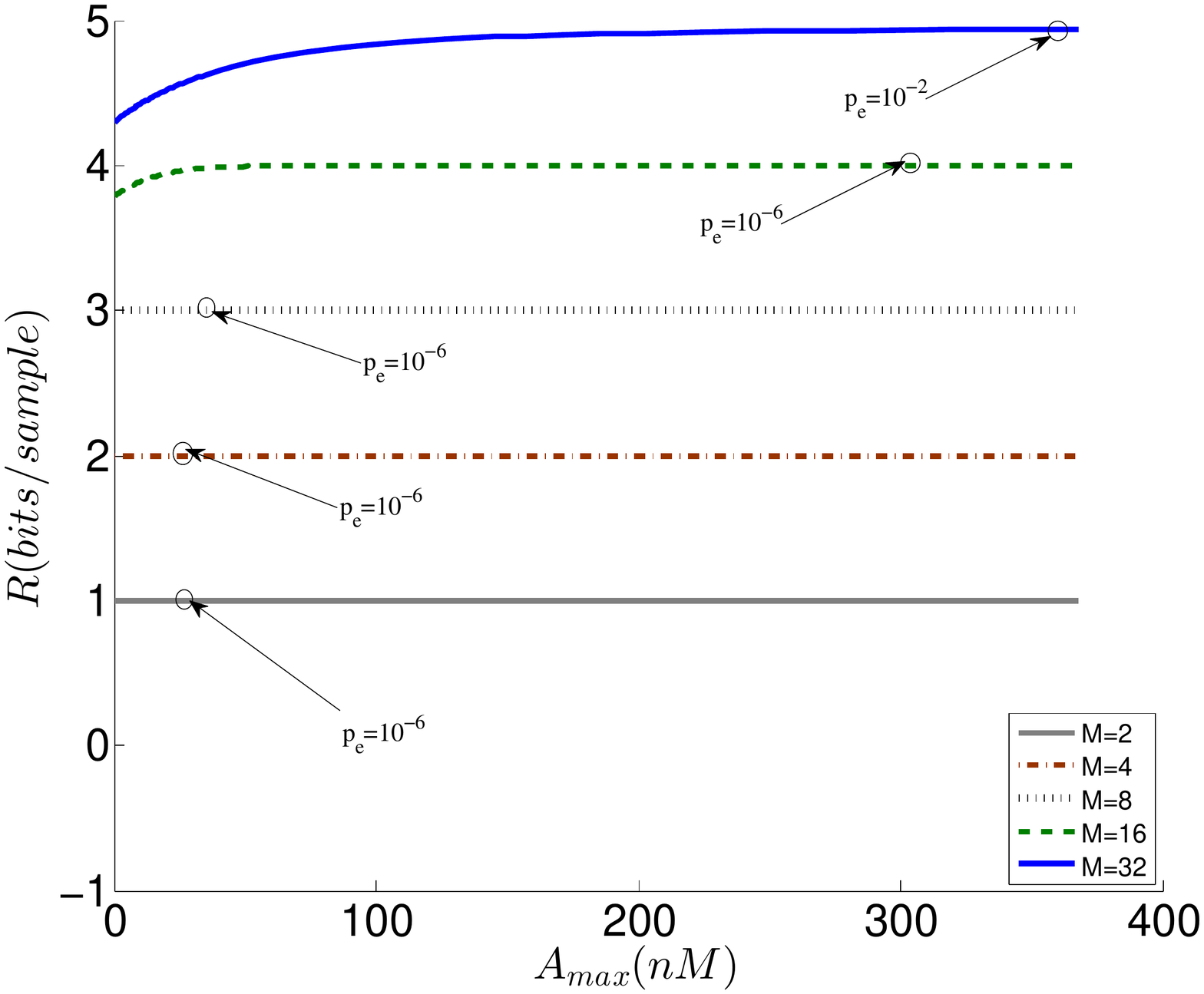}
  \label{fig:modulation}
  }
  \subfigure[The probability of error versus the maximum concentration of molecules at the receiver $A_{max}$ for different M-ary schemes]{
  \includegraphics[width=.48\textwidth]{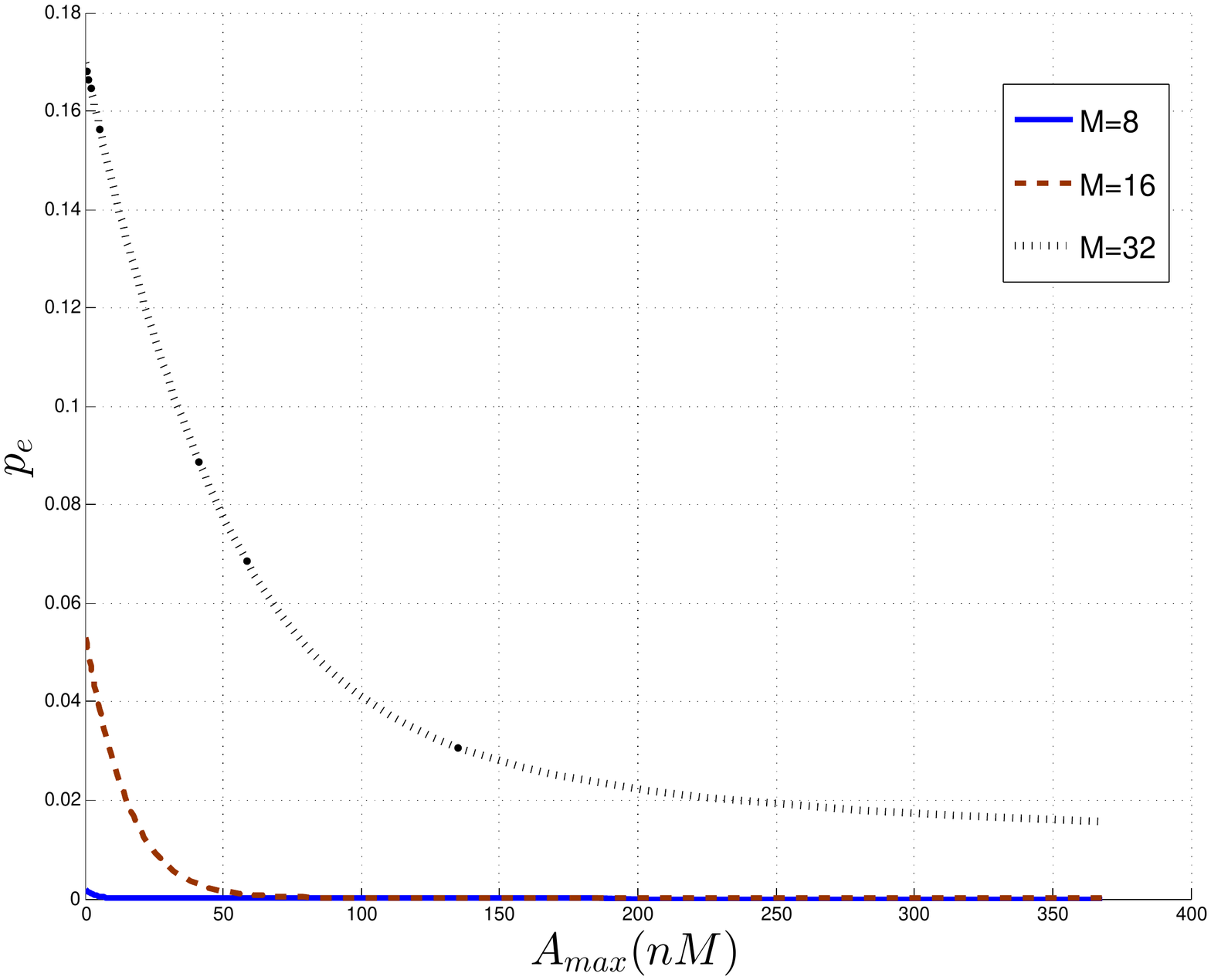}
  \label{fig:modulation_noise}
  }
\vspace{-.1in}
\label{fig:modulation_error}
\caption{Achievable rates and corresponding reliabilities for M-ary schemes}
\end{figure*}
We consider a hard decision scenario in decoding the symbols. In other words, the receiver chooses the closest symbol to the received one. Hence, the error occurs when the detected symbol passes the half way from the previous or the next symbol. As observed in~(\ref{eq:variance_final}), the variance of the noise, and hence $p_{e,i}$ depends on the chosen symbol $i$. Therefore, we have
\begin{equation}
\label{eq:probability_error}
p_{e,i}=1-Pr(\frac{-p_{max}}{2(m-1)} \leq \epsilon_{Yi}  \leq \frac{p_{max}}{2(m-1)}). 
\end{equation}
Here, $\epsilon_{Yi}$ comes from $\mathcal{N}(0,\sigma_i^2)$ where $\sigma_i^2$ can be computed by replacing $p_0$ with $\frac{i}{m-1}p_{max}$ in~(\ref{eq:variance_final}). As discussed in the previous section, variance of the noise is the smallest when the input is closets to $0$ or $1$. Hence, it is intuitive to choose larger weights for the inputs closer to these two points. In our scheme, we use the weights from the optimal distribution calculated by the Blahut-Arimoto algorithm. In Fig.~\ref{fig:modulation}, we have shown the rate of information for different M-ary modulations versus $A_{max}$, the maximum concentration of molecules at the receiver . In this setup, again we have chosen $N=50$ and $\sigma_1^2=0.1$. In addition, the number of bacteria in a node is chosen to be $n=100$.  As shown by the plot, reliable communication (i.e., $p_e=10^{-6}$) is feasible for $M=2, 4, 8, 16$ and the required range is shown as well. For larger number of symbols, reliable communication is not possible as for the case of $M=32$. It was obtained that the least error rate (by maximizing $A_{max}$) would be $10^{-2}$ for $M=32$. This is in contrast with traditional communication in which the reliability of M-ary schemes can be increased bu using more power. Instead, in molecular communication, smaller error rates can be achieved by increasing the number of bacteria $n$.

Finally, in Fig.~{\ref{fig:modulation_noise}}, we have shown the probability of error versus $A_{max}$ for different values of $M$. Note that error for $M=2$ and $4$ is negligible and is not shown in the plot. As we observe in the plot, the probability of error decreases with increasing the range of input (i.e., $A_{max}$) but it is not possible to make the error arbitrarily small for large $M$.

%
%


%

\section{Conclusion}

In this paper, we studied the molecular communication between two populations of bacteria through a diffusion channel. Although, it is aimed at bacteria, the principles developed here can be generalized to analyze the communication of any population of bio entities through molecular signaling in a diffusion channel. The effects of uncertainty in production of molecules, channel parameters and reception process on the overall noise of the communication were considered. We showed that the transmitter noise effect diminishes due to the low-pass property of the reception process. We studied the theoretical limits of the information transfer rate in terms of  number of bacteria per node, noise level and maximum molecule production levels. We observed that the capacity increases with the number of bacteria in the nodes. Finally, we considered M-ary schemes and analyzed the achievable rates and their error probabilities for the special case when $M$ symbols are spaced uniformly. We concluded that for a fixed number of bacteria per node, reliable communication is not possible for large $M$, even with increasing the maximum molecule production at the transmitter node. Instead, reliable communication can be achieved by increasing the number of bacteria in the nodes. Further, we conclude that the information rate of the M-ary scheme is significantly lower than the capacity limit we obtained for the molecular communication. The achievablity of the capacity obtained in this work remains and interesting open problem.
\label{sec:conclusion}

\bibliographystyle{IEEEtran}
\bibliography{Tran_wireless_molecular}
\end{document}